# Observation of the Stimulated Quantum Cherenkov Effect

Saar Nehemia[‡], Raphael Dahan[‡], Michael Shentcis, Ori Reinhardt, Yuval Adiv, Kangpeng Wang, Orr Be'er, Yaniv Kurman, Xihang Shi, Morgan H. Lynch, and Ido Kaminer*

Department of Electrical Engineering, Russell Berrie Nanotechnology Institute, and Solid-State Institute, Technion, Israel Institute of Technology, 32000 Haifa, Israel

[‡] *equal contributors*

* *kaminer@technion.ac.il*

As charged particles surpass the speed of light in an optical medium they produce radiation – analogously to the way jet planes surpass the speed of sound and produce a sonic boom. This radiation emission, known as the Cherenkov effect[1], is among the most fundamental processes in electrodynamics. As such, it is used in numerous applications of particle detectors[2], particle accelerators[3,4], light sources[5], and medical imaging[6]. Surprisingly, all Cherenkov-based applications and experiments thus far were fully described in the framework of *classical* electrodynamics[7,8] even though theoretical work predicts new Cherenkov phenomena coming from *quantum* electrodynamics[9,10]. The quantum description could provide new possibilities for the design of highly controllable light sources and more efficient accelerators and detectors[11-13]. Here, we provide a direct evidence of the quantum nature of the Cherenkov effect and reveal its intrinsic quantum features. By satisfying the Cherenkov condition for the wavefunctions of relativistic electrons and maintaining it over hundreds of microns, *each electron simultaneously* accelerates and decelerates by absorbing and emitting hundreds of photons *in a coherent manner*. We observe this strong interaction in an ultrafast transmission electron microscope, achieving for the first time a **phase-matching between a relativistic electron wavefunction and a propagating light wave**. Consequently, the quantum wavefunction of each electron evolves into a coherent plateau, analogous to a frequency comb in ultrashort laser pulses, containing hundreds of quantized energy peaks. Our findings prove that the delocalized wave nature of electrons can become dominant in *stimulated* interactions. In addition to prospects for known applications of the Cherenkov effect, our work provides a platform for utilizing quantum electrodynamics for new applications in electron microscopy and in free-electron pump-probe spectroscopy[14].

The Cherenkov effect (also called Vavilov-Cherenkov effect) has attracted vast interest since its discovery[1] in 1934 and the Nobel Prize of 1958, yet to this day, all experiments on the subject have been perfectly accounted for by classical electrodynamics. Similarly, all demonstrations of analogous effects in a wide range of fields – such as water waves, acoustics, and even phononics[15]– are also explained entirely classically. This classical Cherenkov effect, and its analogues, enjoy numerous applications[16] in particle identification[2,17,18], medical imaging[6,19], quantum cascade lasers[5], nanophotonics[20-23], and nonlinear optics[24-25]. Likewise, experiments using the **stimulated** Cherenkov effect for electron acceleration and for other electron–laser interactions[3,4,11,26,27] are also described classically.

Interestingly enough, theoretical studies by Ginzburg and Sokolov that date back to 1940 explored the Cherenkov effect within quantum electrodynamics (QED)[9,28]. However, the effects were considered negligible in practice, and the celebrated Lev Landau even said at the time that "quantum corrections are immaterial" in the Cherenkov effect[29]. Despite this critique, recent papers have built on these studies and predicted new intrinsic quantum phenomena in the Cherenkov effect[10,30]. These phenomena led to a far more fundamental puzzle with consequences for the foundations of light-matter interactions: Does general radiation emission depend on the wavefunction shape of the radiating particle or can it always be described in terms of a classical point charge?

In several recent papers, a dependence of both *spontaneous* and *stimulated* radiation on the wavefunction shape was theoretically predicted[13,31-33], despite a recent experiment showing no dependence of *spontaneous* radiation on the wave nature of the electron[34]. Before our work, no experiment has ever shown a dependence of *any radiation phenomena*, especially *stimulated* radiation, on the quantum wave nature of the radiating electron. In this paper, we introduce the first experimental evidence for the dependence of free-electron radiation on the quantum wave nature of the electron, showing that free-electron radiation cannot always be explained by classical point charges (comparison in Fig.1).

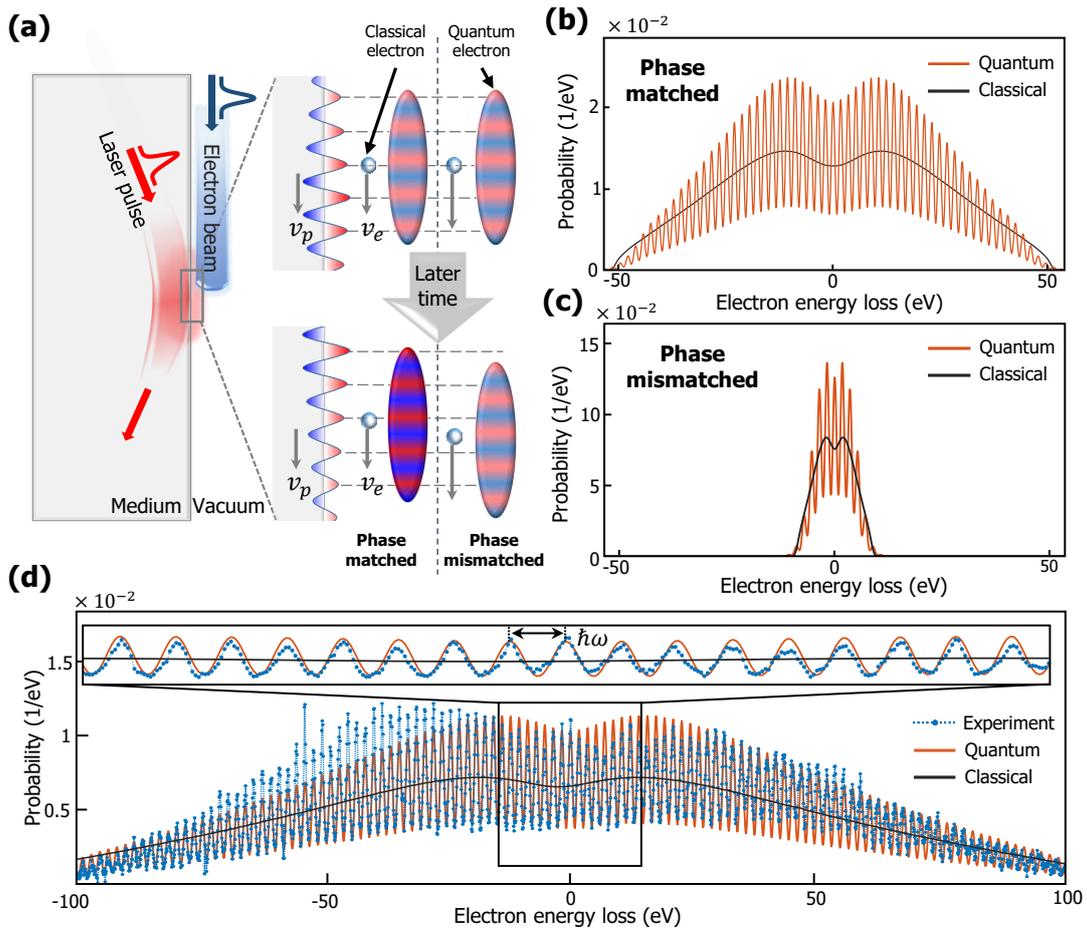

**Figure 1. Quantum vs. classical phase-matching of an electron and light.** Our experimental results show the strong quantum interaction of the electron wavefunction. **(a)** Illustration of the phase-matching effect for both the classical and quantum interpretations. The electron interacts with an evanescent field generated by a laser that is totally internally reflected from an interface. **Classical**: a point electron (small particle) is matched with the light field maximum/minimum and results in accumulated acceleration/deceleration. **Quantum**: an electron wavefunction is matched with multiple cycles of the light field simultaneously and results in a significant modulation of the wavefunction. **(b)–(c)** Comparison of the resulting electron energy spectra. The classical spectrum approximates the average of the quantum spectrum, except for the far edges of the spectrum (more in Supplementary Note 2). **(d)** Example of our measured (blue) and theoretical (orange) electron energy spectrum after a phase-matched interaction. The spectrum shows an energy gain/loss of 100 eV discretized by quanta of photon energy ($\hbar\omega \approx 1.7$ eV). The classical (black) and quantum (orange) calculated spectra are also provided for comparison (details in Supplementary Notes 1 and 2), showing a good fit to the acquired spectrum. Inset: zoom on the range (-15)–15 eV to highlight individual peaks.

Here we observe the quantum stimulated Cherenkov effect and demonstrate the resulting resonant exchange of hundreds of photon quanta with a single electron. By precisely matching the phase velocity of the light wave and the group velocity of the electron wavefunction, we achieve the Cherenkov phase-matching condition, so that each point in the electron wavefunction interacts with a fixed light field direction (Fig. 1a) in a resonant manner. As a result, parts of the electron wavefunction strongly gain energy while other parts strongly lose energy, *simultaneously*. In other words, the same electron **simultaneously absorbs and emits hundreds of photons in our experiment.** The coherent resonant interaction remains constructive over hundreds of microns, resulting in a modulated electron wavefunction that forms a quantized plateau extending over hundreds of electron-volts.

**Results**

The experimental setup that we use to demonstrate the coherent resonant interaction and the resulting Cherenkov phase-matching condition is an ultrafast transmission electron microscope (UTEM)[35-36]. The key to the effects below is our alignment of the electron to graze a surface over hundreds of microns[37] in the UTEM, so that it remains a few hundred nanometers from the surface (Fig. 2). This interaction condition is, to the best of our knowledge, the first realization of such grazing-angle conditions in any transmission electron microscope (see Methods). Using the grazing-angle interaction we maximize the strength of the electron–laser interaction.

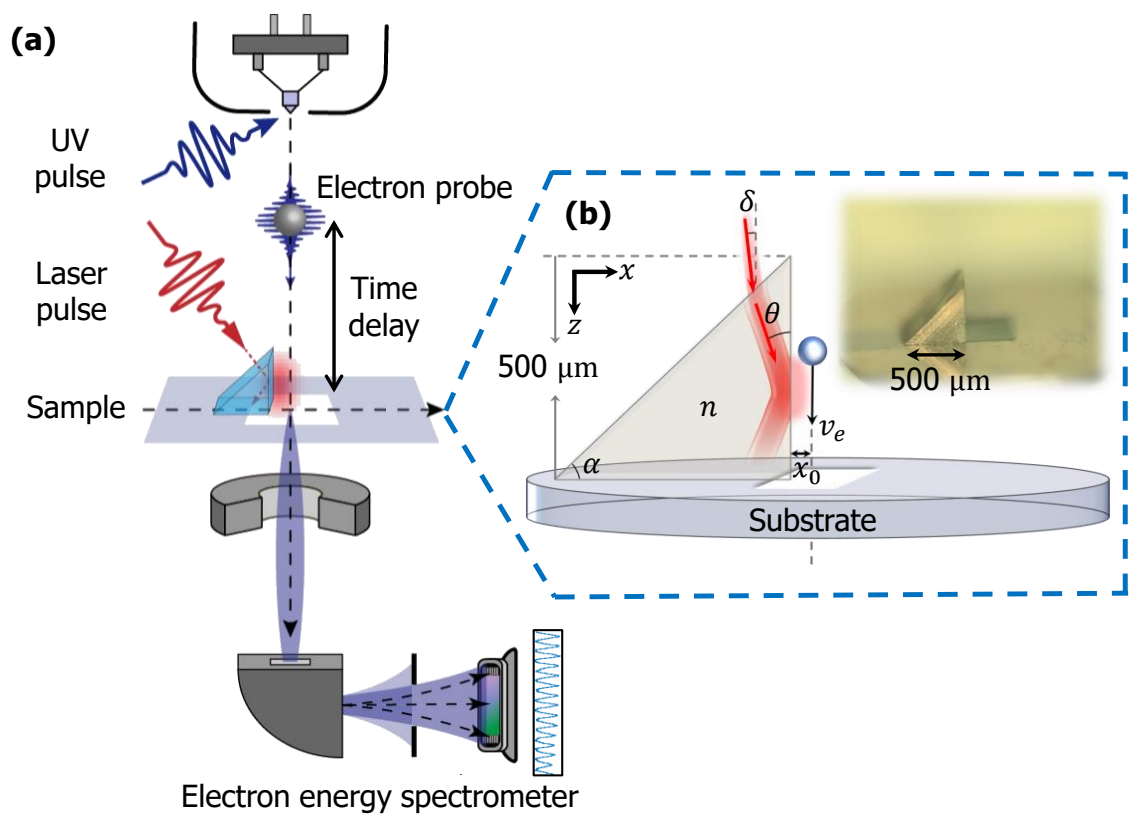

**Figure 2. Experimental setup.** (**a**) Illustration of the ultrafast transmission electron microscope (UTEM) setup, showing the grazing-angle interaction with a prism. The electron pulse is generated by photoexcitation of electrons with a UV pulse. The electrons graze the surface of a prism and interact with an evanescent field generated by another laser pulse that enters the prism and undergoes total internal reflection from the same surface. We measure the electrons with an electron energy spectrometer. (**b**) Zoom-in on the interaction area along the prism surface. Inset: an image of the prism positioned on the edge of a hole through which the electrons pass (parameters defined in Methods).

The UTEM setup was previously used[38-40] to study the quantized interaction of free electrons with laser excitations that were localized or propagating transverse to the electron velocity (e.g., Refs. 41-47). Such interactions were used to demonstrate free-electron Rabi oscillations and quantum walks[41], particle–wave duality with single electrons[42], laser-controlled electron angular momentum[45], and measurement of the lifetime of a cavity photon[48]. In contrast, our work demonstrates that by coupling an electron to a light wave propagating *parallel* to the electron's velocity, and achieving phase-matching of the electron with the light wave, we can observe quantized stimulated *radiation* effects such as the stimulated Cherenkov effect.

In all previous quantized electron-laser interactions, including our work, the interaction strength can be quantified by a single dimensionless coupling constant $g$[39,40]. This coupling constant is derived by integration of the electric field along the electron trajectory (at coordinates $(x_0, y_0)$ and along the $\hat{z}$ axis):

$$g(x_0, y_0) = \frac{q_e}{\hbar\omega} \int_{-\infty}^{\infty} E_z(x_0, y_0, z) e^{-i\omega z/v_e} dz, \qquad (1)$$

where $E_z$ is the phasor of the electric field $z$ component, $q_e$ is the electron charge, $\omega$ is the laser frequency, and $v_e$ is the electron's velocity. In our setup, the electron interacts with an evanescent wave that is expressed as $E_z(\mathbf{r}) = E_{0,z} e^{i\mathbf{k}\cdot\mathbf{r}} = E_{0,z} e^{-K_x x + i k_z z}$ ($K_x > 0, x > 0$), where $E_{0,z}$ and $\mathbf{k} = (k_x = iK_x, k_y = 0, k_z)$ are the amplitude and wavevector of the evanescent electric field, respectively. Substituting in Eq. 1, The coupling constant $g$ is given by

$$g(x_0) = \frac{q_e}{\hbar\omega} \int_{-\infty}^{\infty} E_{0,z} e^{-K_x x_0 + i(k_z - \omega/v_e)z} dz. \qquad (2)$$

The difference $k_z - \omega/v_e$ in the exponent is the phase-mismatch between the electron and the field.

The key to realizing the Cherenkov effect is to satisfy the phase-matching condition ($v_e = \omega/k_z$), which for a long interaction length also results in a coupling constant that grows macroscopically with the system size[49]. Quantitively, we can define an *effective interaction length* $L_{\text{eff}}$ such that $|g| = \frac{q_e E_{0,z} L_{\text{eff}}}{\hbar\omega}$. In previous experiments (e.g., Refs. 41-47, 50, 51) the effective length of interaction is roughly $L_{\text{eff}} \lesssim \lambda$ (several hundred *nanometers*). However, we were able to increase it to $L_{\text{eff}} \gg \lambda$ (several hundred *microns*), using our grazing-angle setup[37,52]. As a result, we achieved an interaction with a coupling constant $g$ much larger than previous works.

To achieve a strong interaction in our experiment, we satisfy the phase-matching condition by reducing the phase velocity of light using a dielectric medium – a glass prism. As the light is totally internally reflected inside the prism, an evanescent tail extends outside the medium into vacuum, where it interacts with the electron that grazes the surface along several hundred microns (Figs. 1a and 2). This grazing interaction is sometimes called the Cherenkov–Landau effect[53]. A similar prism setup was used previously[4] to achieve the phase-matching condition in *classical electrodynamics* using a scanning electron microscope. We demonstrate the *quantum* analogue of the phase-matching condition for the first time by using a coherent electron wavefunction and measure it with an energy resolution better than $\hbar\omega$, thereby revealing the quantum features of the interaction.

Looking at the quantum description of the interaction, the phase-matching condition from Eq. 2 ($v_e = \omega/k_z$) is reduced to the well-known Cherenkov condition: $n\cos\theta = c/v_e$, with $n$ being the prism's index of refraction and $\theta$ is the angle between the electron and the refracted light inside the prism. See Supplementary Note 3 for details.

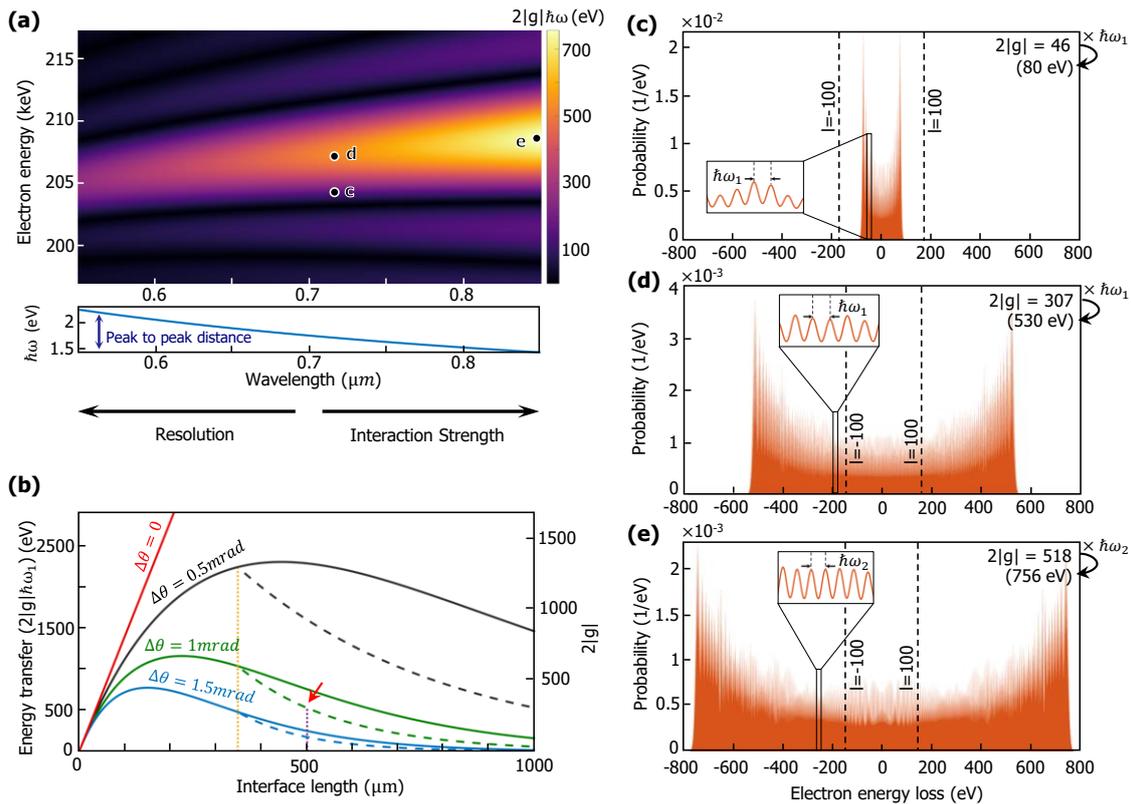

**Figure 3. Conditions for phase-matching: theoretical analysis. (a)** The interaction strength quantified by the energy spread $2|g|\hbar\omega$ as a function of the electron kinetic energy and the wavelength of the laser, showing the curve of perfect phase-matching. Shallow panel below the map: Reminder of the photon energy vs. wavelength to emphasize that the distance between peaks in the spectra changes with the wavelength. As we increase the wavelength, we obtain a stronger interaction but gradually lose the ability to resolve individual peaks. **(b)** The interaction strength as a function of the prism length for different divergence angles of the electron beam ($\Delta\theta = 0$ is a perfectly collimated beam). By fitting our data, we estimate a beam divergence of $\Delta\theta \approx 1\,\text{mrad} = 0.057°$. This parameter also includes a deviation from a perfect straight trajectory, because the beam follows a helical trajectory created by the magnetic field of the objective lens. This divergence keeps the electron beam farther away from the prism for larger interface lengths, which explains the eventual decay of all curves except for the perfectly collimated beam. A threshold length is set by the laser spot size projected on the interface (~350 μm, yellow dotted line), above which an increase in the length of the interface results only in a decay in $g$ (dashed curves). Red arrow: The point at which we work in our experiment. **(c)–(e)** Examples of electron energy spectra calculated with different coupling constants showing weak ($E_e = 204.5$ eV, $\lambda_1 = 730$ nm (1.7 eV)), intermediate ($E_e = 207.2$ eV, $\lambda_1 = 730$ nm (1.7 eV)), and strong ($E_e = 208.5$ eV, $\lambda_2 = 850$ nm (1.46 eV)) interaction strengths, respectively. The calculations for this figure use our experimental parameters unless stated otherwise (see Methods for more information).

Figure 3 analyses the effect of different parameters on the phase-matching and emphasizes **the sensitivity of the strength of the interaction to the electron energy.** The acceleration voltage controls the electron kinetic energy ($E_e$ =207.2 keV) and determines its velocity ($v_e$ =0.7027c). Because of the material dispersion, the refractive index of prism changes with the laser wavelength e.g., n=1.512 at $\lambda$ =730 nm. The strongest interaction at each wavelength in Fig. 3a follows a curve that satisfies the phase-matching condition. A small change, of only 2 keV (<1%), in the electron energy can result in $g$ changing by an order of magnitude. The electron energy spectra in Figs. 3(c)–(e) highlight the importance of precise phase-matching for the interaction strength: roughly, the maximum number of photons exchanged is $2|g|$, and the energy spread (the edge of each spectrum) is $2\hbar\omega|g|$. The theoretical analysis is provided in Supplementary Note 1.

An optimal interaction has to balance important trade-offs: The interaction is indeed stronger for longer wavelengths (Fig. 3a) but the quantum features are harder to measure because the distance between adjacent peaks shrinks (insets in Figs. 3(c)–(e)). Another tradeoff is that a longer medium interface increases $g$ but also requires that the electron stays farther away from the interface because of its unavoidable spread angle (Fig. 3b), which creates an exponential decrease in the interaction strength (dashed lines in Fig. 3b)[52].

Figure 4 shows several measurements of strong interactions with high $g$ values that resulted from the Cherenkov phase-matched interaction. In the blue curve in Fig. 4a we achieved $|g|{\sim}150$ that matches a maximum energy gain/loss of 510 eV, also corresponding to the zoom-in panels in Fig. 4b and the comparison with theory in Fig. 4c. By using a shorter duration and a more intense laser pulse, we achieved $|g| > 250$ that matches a maximum energy gain/loss $> 850$ eV (pink energy spectra in Fig. 4a). However, this interaction involves only part of the electron distribution, leaving a large near-zero peak in the electron energy spectrum. The result is a free-electron comb where the electron exchanges hundreds of photons with the field, becoming a coherent superposition of energy peaks in the form of a quantized plateau (Fig. 4b).

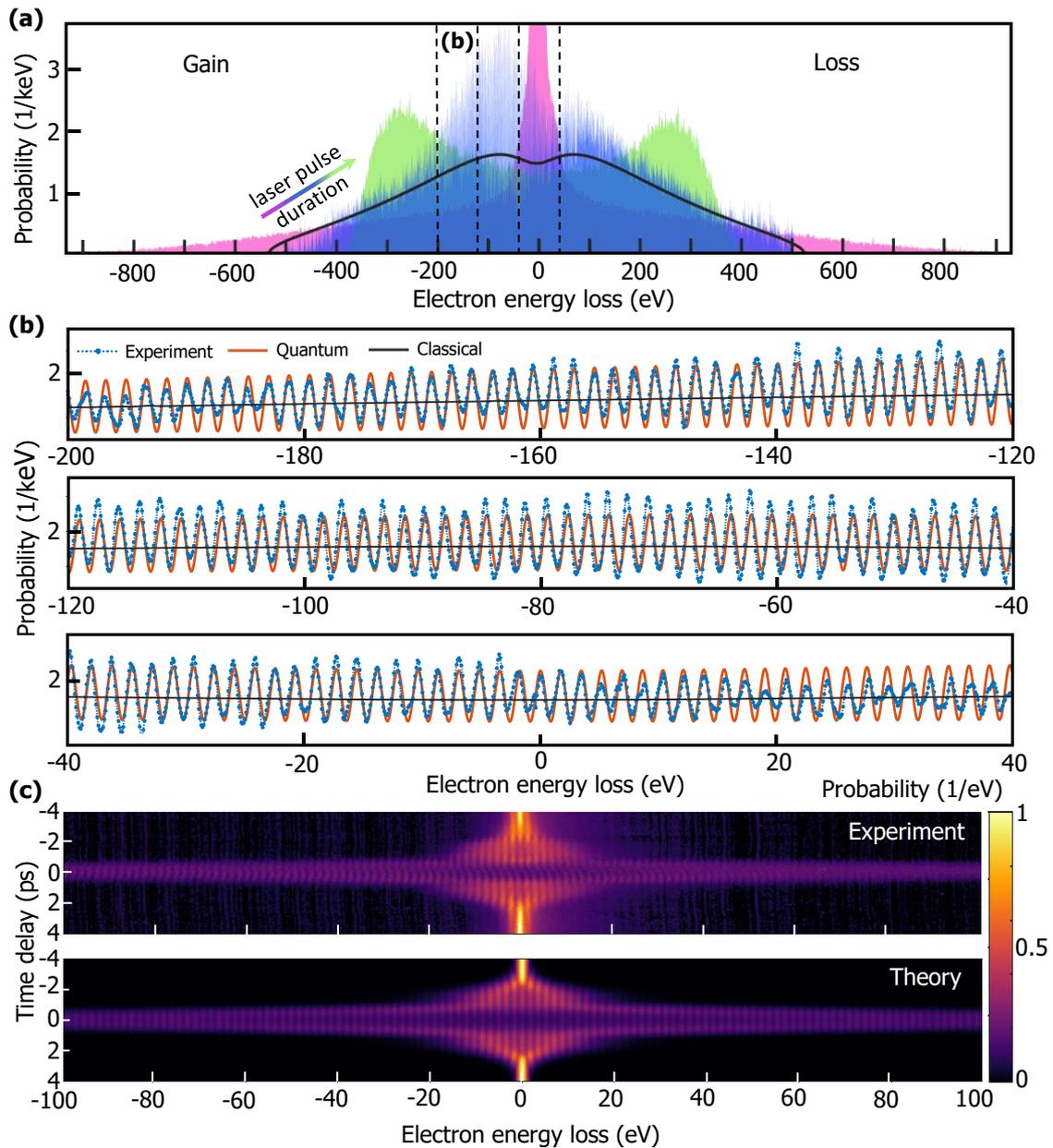

**Figure 4. Experimental results: Coherent electron energy comb in a plateau. Record strong interactions as a function of the time delay between the electron and the laser pulses.** (a) Acquired electron energy spectra for different laser pulse durations (pink- 280 fs, blue- 600 fs, green- 1300 fs), while the electron's pulse duration is kept at 300 fs. As we increase the laser pulse duration, the entire electron distribution feels the same laser amplitude, which creates a double-peaked structure in the electron energy spectrum (green). As we decrease the laser pulse duration and thereby increase its peak intensity, the maximal energy transfer increases to >850 electron-volts (pink), and yet, only part of the electron distribution feels the maximal laser field. (b) Zoom-in of the blue spectrum in (a), with comparison to the quantum theory and classical theory (described in Supplementary Notes 1 and 2). We were able to resolve **individual quantized peaks over a range of hundreds of electron-volts** by shifting the range of the spectrometer and **collecting the entire spectrum slice by slice**. (c) Experimental and theoretical time delay scans, varying the delay between the laser and electron pulses. We obtain a good match between theory and experiment by modeling the laser's temporal pulse shape to incorporate changes caused by our optical parametric amplifier[54] and by considering the effective interaction length (Supplementary Note 4b). All the results in this work are presented at the point of maximal interaction (time-zero).

Importantly, Fig. 4 shows that the quantization of the electron energy spectrum in the form of a comb can spread over more than a thousand electron-volts. An electron energy comb has been previously observed only in pulsed photoexcitation of *bound electrons*, as in the phenomenon of above-threshold ionization[55]. Such spectral features in electrons are the reason for the production of a similar comb of high laser harmonics, which opened the field of attosecond science[56]. All these rely on initially bound electrons that absorb multiple photons, but never on *free electrons*. Free electrons have markedly different physical phenomena and applications as compared to bound electrons. For example, unlike bound electrons, free electrons can have relativistic velocities and therefore induce new relativistic effects, such as the Cherenkov effect that we observed in this work.

**Discussion**

Our work experimentally demonstrates the quantum nature of the stimulated Cherenkov effect and in general of any stimulated radiation processes by a free electron. We find quantum features that arise from the electron being a wavefunction instead of just a point particle. Our results thus conform to theoretical predictions about the dependence of *stimulated* radiation on the quantum wavefunction of the emitting electron[13,33]. Nevertheless, the prospects and mere possibility of *spontaneous* radiation depending on the wavefunction still require further research: one experiment on free-electron spontaneous radiation showed no wavefunction-dependence[34], in contrast to another experiment on spontaneous nearfield excitations (rather than radiation) that did show such dependence[57].

We find a good match between theoretical prediction and experimental results of the electron energy spectrum (Fig. 4c). Yet in some cases, as in Fig. 1d, we notice an asymmetry between the gain and loss sides in our measured spectrum. This asymmetry can be explained by several energy loss mechanisms that differ from the stimulated Cherenkov effect, such as bulk plasmon emission and core losses. Another potential reason for the asymmetry is more intriguing: the overall strength and length of the interaction, producing 100's of energy peaks over a distance of 100's of microns, involve corrections due to the electron dispersion that can no longer be neglected[39,40] as in all UTEM experiments so far.

The phase-matching we obtain from the Cherenkov condition is analogous to the phase-matching utilized in dielectric laser acceleration (DLA)[58-60], where a tailor-designed laser-driven nanophotonic structure accelerates the particles. In our case, it is the evanescent mode propagating along the surface of the prism that acts as the effective means of acceleration. Thus, a UTEM system can complement the existing experimental setups used for testing DLA devices (our system operates at 40–200 keV). Most importantly, the wave nature of the electron provides an additional degree of control to the acceleration process, thereby opening up another avenue of research on these systems. The comparison of the quantum theory with the conventional classical one shows additional fine details (individual energy peaks) and **provides a more accurate prediction of the regimes of highest acceleration and deceleration** (Supplementary Note 2). Consequently, it is interesting to explore further implications of the quantum nature of the electron for the design of future DLAs.

More generally, analogous phase-matching conditions appear in other free-electron radiation effects such as the (stimulated-)Smith–Purcell effect[61-63] and various undulator concepts[7,8]. In all experimental work on these effects, the electron has always been considered to be a classical point charge. We now show a completely new regime where these kinds of effects are essentially quantum and require the electron to be a wavefunction to correctly explain the experiment. We expect analogous experiments to our work here to reveal underlying quantum wave effects and quantized electron energy exchanges *in all these systems*.

The phase-matching interaction in the UTEM opens the door for the exploration of several new phenomena. In addition to the stimulated Cherenkov radiation in our experiment, the electron is also expected to emit spontaneous Cherenkov radiation throughout its motion. However, since the electron has been modulated into a comb of energy peaks, the subsequent Cherenkov emission may also be composed of multiple radiation orders at different angles and frequencies[64]. Another intriguing phenomenon that could be considered is ultrastrong coupling in electron–photon interactions[30]. It was predicted recently[65] that combining a high-Q photonic cavity with a phase-matched interaction could lead to an efficient single-electron–single-photon interaction[65,66]. With this goal in mind, we have recently demonstrated the stimulated free-electron interaction with photonic cavities[48], which can serve as the platform for free-electron cavity QED interactions in the UTEM. Looking ahead, we envision combining the phase-matched interaction with a photonic cavity as a route to achieving ultrastrong coupling of free electrons and light. The cavity will channel emitted photons that can then be resonantly reabsorbed by the electron, creating a strongly-coupled electron–photon hybrid. This hybrid will enable the exploration of novel processes such as free-electron Lamb shifts, extreme mass renormalizations, and potentially even cavity-mediated Cooper pairs of free electrons.

## Methods

*Experimental setup: ultrafast transmission electron microscope (UTEM)*

All the experiments presented in this work were conducted using a UTEM (Jeol-2100 Plus) in nano-beam diffraction (NBD) mode operating at $E_e \approx 207.2 \text{KeV}$. The setup consisted of a right-angle prism made of BK7 (index of refraction: $n = 1.512 @ \lambda = 730$ nm) at a height of 500 μm. The prism was placed on a specially designed TEM holder with one of its faces parallel to the electron beam. By splitting the laser source (LightConversion, Carbide), we created a pump-probe setup, where one pulse is converted to a UV pulse to generate photo-electrons (probe) and the second pulse is converted to visible light, which excites the sample to create the desired EM field (pump). A relative delay between the pump and probe pulses gives us precise control over the relative arrival time of the electron and laser pulses, which in our experiment also describes the location of their interaction. The pump pulse (730 nm) is coupled into the prism and undergoes total internal reflection from the surface of the prism (parallel to the electron beam), exciting an evanescent nearfield that interacts with the electrons grazing the same surface (Fig. 2). We chose a wavelength of 730 nm, considering our optical parametric amplifier conversion efficiency, while limiting ourselves to a range where we maintain the ability to resolve individual peaks (our zero-loss peak width is ~1.1 eV in all figures except for Fig. 1d where it is ~0.6 eV).

We use a Gatan electron energy loss spectrometer (EELS) with a resolution of ~0.1 eV, allowing us to reveal the hidden quantum features of the interaction. The actual resolution limit for individual energy peaks is the width of the electron zero-loss peak given above. In Fig. 3, we show three examples of EELS of the stimulated Cherenkov effect. We succeeded in observing electrons that gain or lose up to 300 quanta of energy with high energy resolution – identifying the individual peaks by recording several energy slices at different shifts.

To determine the correct parameters for the phase-matching, we calculated the beam path inside the prism (see Supplementary Note 4b). This calculation determines the required angle of incidence of light before its transmission into the prism (40.0°), which is the prism's base angle (45.0°) minus the laser coupling angle (5.0°). This incidence angle yields the Cherenkov angle (70.2°) of the refracted light relative to the surface of the prism for the chosen electron's kinetic energy (207.2 keV) and laser's wavelength (730 nm).

*Grazing-angle interaction: alignment challenges*

The main experimental challenge of this work was the alignment of the electron beam to graze the prism's surface and to interact with the evanescent laser field near the prism surface. Any small tilt of the beam relative to the prism results in the electron beam's trajectory being pushed farther away from the surface of the prism, weakening the interaction significantly. The electron beam size also controls the spatial overlap (see Supplementary Note 4a) that results in a more "rounded edge" spectrum when using a bigger electron beam diameter (controlled by the condenser aperture).

To achieve parallel electron illumination, we chose to work in nano-beam diffraction (NBD) mode with a 70 μm condenser aperture. Then, the current center is set by wobbling the objective's current (first with no condenser aperture to obtain sufficient counts) while minimizing the spot movement (spot alignment together with beam tilt). The prism tilt angle is set by minimizing the prism shadow. Then, the condenser aperture is inserted (70 μm diameter) to obtain a smaller spot size and the same steps are repeated. The estimated convergence angle of the electron beam is 1 mrad, which translates to an average distance from the prism of $x_0 \sim 250$ nm (see Supplementary Note 4a).

As a final step, we minimize the deviation of the electron motion from a parallel path. The electron always follows a slightly helical path that arises from the strong magnetic field in the objective lens. We adjust the beam tilt while looking at the change in the prism's shadow while wobbling the objective. We estimate the helix radius and pitch using the Lorentz force $\boldsymbol{F} = q_e \boldsymbol{v_e} \times \boldsymbol{B}$ for our magnetic field of 1.4 T and electron convergence angle of (worst case scenario) 1 mrad relative to the objective axis. We obtained a helical path with a pitch of 5.38 mm and radius of 0.86 microns, which changes the beam distance from our prism by 100 nm. Additionally, it is important to note that because our sample is considerably taller than regular samples, our interaction may be affected by the inhomogeneity of the magnetic field near the pole pieces.

Fig. 3a shows that the interaction strength as a function of the acceleration voltage has multiple sidelobes when the interaction is truncated by the prism surface length (500 μm). The distance between the sidelobes scales inversely with the interaction length; see Supplementary Note 3. These sidelobes disappear when the transverse spatial Gaussian shape of the pump laser is considered (see Supplementary Note 4b Fig. 4S). The laser spot size on the surface of the prism was on the order of ~350 μm for an incident pump laser spot size of 100 μm. Taking these parameters into account, we arrive at a general formula for the electron energy spectrum, with which we fit the time scan data (Fig. 4c and Supplementary Note 4b).

## Acknowledgements

We thank the IDES Company and especially Dr. Sang Tae Park for illuminating discussions and advice. The experiments were performed on the UTEM of the I. K. AdQuanta group installed in the electron microscopy center (MIKA) in the Department of Material Science and Engineering at the Technion.

Therefore, it seems that $g$ would diverge for an infinite interaction length, but in practice, it is bounded by the finite length of the interaction[52].